\documentclass[aps,prl,twocolumn,superscriptaddress,showpacs,preprintnumbers,amsmath,amssymb,amstex,amsfonts,tightenlines]{revtex4}

\usepackage{graphicx} % Include figure files
\usepackage{dcolumn}  % Align table columns on decimal point
\usepackage{xspace}

\newcommand{\invfb}{\ensuremath{\mathrm{fb}^{-1}}\xspace}
\newcommand{\dm}{\ensuremath{\Delta m_{d}}\xspace}
\newcommand{\dz}{\ensuremath{\Delta z}\xspace}
\newcommand{\dt}{\ensuremath{\Delta t}\xspace}
\newcommand{\fpfz}{\ensuremath{f_{+}/\hspace*{-0.2ex}f_{0}}\xspace}
\newcommand{\ups}{\ensuremath{\Upsilon (4S)}\xspace}
\newcommand{\rcos}{\ensuremath{\Re (\cos \theta)}\xspace}
\newcommand{\icos}{\ensuremath{\Im (\cos \theta)}\xspace}
\newcommand{\B}{\ensuremath{B}\xspace}
\newcommand{\jpsi}{\ensuremath{J\hspace*{-0.5ex}/\hspace*{-0.2ex}\psi}\xspace}
\newcommand{\CPT}{\emph{CPT}\xspace}
\newcommand{\CP}{\emph{CP}\xspace}

\newcommand{\mev}{\ensuremath{\mathrm{MeV}}\xspace}

\newcommand{\gevc}{\ensuremath{\mathrm{GeV}\hspace*{-0.5ex}/c}\xspace}
\newcommand{\mevcc}{\ensuremath{\mathrm{MeV}\hspace*{-0.5ex}/c^2}\xspace}
\newcommand{\gevcc}{\ensuremath{\mathrm{GeV}\hspace*{-0.5ex}/c^2}\xspace}
\newcommand{\invps}{\ensuremath{\mathrm{ps}^{-1}}\xspace}
\newcommand{\dedx}{\ensuremath{dE\hspace*{-0.3ex}/\hspace*{-0.3ex}dx}\xspace}

\graphicspath{{./}{eps/}}

\begin{document}

\title{Studies of $\mathbf{B^0}$-$\mathbf{\bar{B}^0}$
  Mixing Properties with Inclusive Dilepton Events}

\affiliation{Budker Institute of Nuclear Physics, Novosibirsk}
\affiliation{Chiba University, Chiba}
\affiliation{Chuo University, Tokyo}
\affiliation{University of Cincinnati, Cincinnati, Ohio 45221}
\affiliation{University of Frankfurt, Frankfurt}
\affiliation{Gyeongsang National University, Chinju}
\affiliation{University of Hawaii, Honolulu, Hawaii 96822}
\affiliation{High Energy Accelerator Research Organization (KEK), Tsukuba}
\affiliation{Hiroshima Institute of Technology, Hiroshima}
\affiliation{Institute of High Energy Physics, Chinese Academy of Sciences, Beijing}
\affiliation{Institute of High Energy Physics, Vienna}
\affiliation{Institute for Theoretical and Experimental Physics, Moscow}
\affiliation{J. Stefan Institute, Ljubljana}
\affiliation{Kanagawa University, Yokohama}
\affiliation{Korea University, Seoul}
\affiliation{Kyoto University, Kyoto}
\affiliation{Kyungpook National University, Taegu}
\affiliation{Institut de Physique des Hautes \'Energies, Universit\'e de Lausanne, Lausanne}
\affiliation{University of Ljubljana, Ljubljana}
\affiliation{University of Maribor, Maribor}
\affiliation{University of Melbourne, Victoria}
\affiliation{Nagoya University, Nagoya}
\affiliation{Nara Women's University, Nara}
\affiliation{National Lien-Ho Institute of Technology, Miao Li}
\affiliation{National Taiwan University, Taipei}
\affiliation{H. Niewodniczanski Institute of Nuclear Physics, Krakow}
\affiliation{Nihon Dental College, Niigata}
\affiliation{Niigata University, Niigata}
\affiliation{Osaka City University, Osaka}
\affiliation{Osaka University, Osaka}
\affiliation{Panjab University, Chandigarh}
\affiliation{Peking University, Beijing}
\affiliation{Princeton University, Princeton, New Jersey 08545}
\affiliation{RIKEN BNL Research Center, Upton, New York 11973}
\affiliation{Saga University, Saga}
\affiliation{University of Science and Technology of China, Hefei}
\affiliation{Seoul National University, Seoul}
\affiliation{Sungkyunkwan University, Suwon}
\affiliation{University of Sydney, Sydney NSW}
\affiliation{Tata Institute of Fundamental Research, Bombay}
\affiliation{Toho University, Funabashi}
\affiliation{Tohoku Gakuin University, Tagajo}
\affiliation{Tohoku University, Sendai}
\affiliation{University of Tokyo, Tokyo}
\affiliation{Tokyo Institute of Technology, Tokyo}
\affiliation{Tokyo Metropolitan University, Tokyo}
\affiliation{Tokyo University of Agriculture and Technology, Tokyo}
\affiliation{Toyama National College of Maritime Technology, Toyama}
\affiliation{University of Tsukuba, Tsukuba}
\affiliation{Utkal University, Bhubaneswer}
\affiliation{Virginia Polytechnic Institute and State University, Blacksburg, Virginia 24061}
\affiliation{Yokkaichi University, Yokkaichi}
\affiliation{Yonsei University, Seoul}
  \author{N.~C.~Hastings}\affiliation{University of Melbourne, Victoria} % Melbourne
  \author{K.~Abe}\affiliation{High Energy Accelerator Research Organization (KEK), Tsukuba} % KEK
  \author{K.~Abe}\affiliation{Tohoku Gakuin University, Tagajo} % TohokuGakuin
  \author{R.~Abe}\affiliation{Niigata University, Niigata} % Niigata
  \author{T.~Abe}\affiliation{Tohoku University, Sendai} % Tohoku
  \author{I.~Adachi}\affiliation{High Energy Accelerator Research Organization (KEK), Tsukuba} % KEK
  \author{H.~Aihara}\affiliation{University of Tokyo, Tokyo} % Tokyo
  \author{M.~Akatsu}\affiliation{Nagoya University, Nagoya} % Nagoya
  \author{Y.~Asano}\affiliation{University of Tsukuba, Tsukuba} % Tsukuba
  \author{T.~Aso}\affiliation{Toyama National College of Maritime Technology, Toyama} % Toyama
% ...
% Ljubljana
\collaboration{The Belle Collaboration}

\begin{abstract}
  We report a precise determination of the $B^0$-$\bar B^0$ mixing
  parameter \dm based on the time evolution of same-sign and
  opposite-sign dilepton yields in \ups decays. Data were collected
  with the Belle detector at KEKB. Using data samples of 29.4 \invfb
  recorded at the \ups resonance and 3.0 \invfb recorded at an energy
  60~MeV below the resonance, we measure
  $\dm~\!=~\!(0.503~\!\pm~\!0.008(\mathrm{stat})~\!\pm~\!0.010(\mathrm{sys}))~\invps$.
  From the same analysis, we also measure the ratio of charged and
  neutral \B meson production at the \ups,
  $\fpfz~\!=~\!1.01~\!\pm~\!0.03(\mathrm{stat}) \pm
  0.09(\mathrm{sys})$, and \CPT violation parameters in $B^0$-$\bar
  B^0$ mixing,
  $\rcos~\!=~\!0.00~\!\pm~\!0.12(\mathrm{stat})~\!\pm~\!0.01(\mathrm{sys})$
  and
  $\icos~\!=~\!0.03~\!\pm~\!0.01(\mathrm{stat})~\!\pm~\!0.03(\mathrm{sys})$.
\end{abstract}


\maketitle

{\renewcommand{\thefootnote}{\fnsymbol{footnote}}}
\setcounter{footnote}{0}

\section{Introduction}
The mass difference of the $B^0$-$\bar{B}^0$ mass eigen states, \dm is
a fundamental parameter in the \B meson system. The techniques that
have been employed to measure it in $\ups \rightarrow B^0 \bar{B}^0$
decays fall into two categories, namely inclusive and exclusive. Among
the inclusive methods, the analysis of events where both \B mesons
decay into a final state that includes a high momentum lepton provides
the largest event sample~\cite{PRLDL} and is well suited for further
high precision measurements of the time evolution in the $B^0
\bar{B}^0$ system. This system exhibits sensitivity not only to \dm,
but also to other potentially interesting phenomena such as \CP
violation in mixing, the decay width difference between the two mass
eigenstates and possible \CPT violation in mixing~\cite{PRD}.

Without the assumption of \CPT invariance, the flavour and  
mass eigenstates of the neutral \B mesons are related by 
\begin{eqnarray}
| B_{H} \rangle &=& p \; | B^{0} \rangle + q \; |\bar{B}^0 \rangle \nonumber \\
| B_{L\:} \rangle &=& p^\prime  | B^{0} \rangle - q^\prime|\bar{B}^0 \rangle.
\end{eqnarray}
The coefficients $p$, $q$, $p^\prime$ and $q^\prime$ can be expressed
in terms of the complex parameters $\theta$ and $\phi$ by
$\frac{q}{p}=\tan(\frac{\theta}{2})e^{i\phi}$ and
$\frac{q^\prime}{p^\prime}=\cot(\frac{\theta}{2})e^{i\phi}$. \CP is
violated if $\Im(\phi) \neq 0$, and \CPT is violated if $\theta \neq
\frac{\pi}{2}$~\cite{PRD}. The time-dependent decay rates are given
by~\cite{PRD}
\begin{eqnarray}
\lefteqn{\Gamma_{\ups \rightarrow \ell^{\pm}\ell^{\pm}}(\dt) =
 \frac{|A_{\ell}|^{4}}{8\tau_{B^0}} e^{-|\dt|/\tau_{B^0}} \times
} \nonumber \\
&& | \sin\theta e^{\mp i\phi} |^{2}   
\left[ \cosh\left(\textstyle{ \frac{\Delta \Gamma}{2}} \dt\right)
  - \cos\left(\dm \dt\right) \right]
\label{eq:noassumption_ss}
\end{eqnarray}
for same-sign dilepton (SS) events, and
\begin{eqnarray}
  &\!\!\!&\!\!\!
  \Gamma_{\ups \rightarrow \ell^{+}\ell^{-}}(\dt) =
  \frac{{|A_{\ell}|}^4}{4\tau_{B^0}}e^{-|\dt|/\tau_{B^0}}
  \times
  \nonumber \\
  &\!\!\!&\!\!\!
  \{(1+|\cos\theta|^2)
    \cosh\left(\textstyle{\frac{\Delta\Gamma}{2}}\dt\right) 
  + (1-|\cos\theta|^2)  \cos(\dm \dt) 
  \nonumber  \\
  &\!\!\!&\!\!\!
  + \; 2\Re(\cos\theta)\sinh\left(\textstyle{\frac{\Delta\Gamma}{2}}
  \dt \right)
  - 2\Im(\cos\theta)\sin(\Delta m_d \dt) \}
\label{eq:noassumption_os}
\end{eqnarray}
for opposite-sign (OS) events.
Here, we assume \CP is conserved in the flavour specific semileptonic
decay amplitudes of the neutral \B mesons and set $A_{\ell} \equiv
\langle X^- \ell^+ \nu_\ell | B^0 \rangle$ and $\bar{A_{\ell}} \equiv
\langle X^+ \ell^- \bar\nu_\ell | \bar{B}^0 \rangle$ to be equal.
$\Delta m_d$ and $\Delta \Gamma$ are the differences in the mass and
decay width between the two mass eigenstates of the neutral \B meson,
$\Gamma=1/\tau_{B^0}$ is the average decay width of the two mass
eigenstates, \dt is the proper time difference between the two
\B meson decays and is defined as $\dt \equiv t(\ell^+) -
t(\ell^-)$ for the OS events, while the absolute value is taken for SS
events.

In equation~\ref{eq:noassumption_ss}, \CP violation appears as a
difference in the $\ell^+\ell^+$ and $\ell^-\ell^-$ rates, in case of
a non-zero $\Im(\phi)$. It does not depend on $\theta$, $\Delta\Gamma$
or the fraction of mixed events. The last two terms in
equation~\ref{eq:noassumption_os} are clearly asymmetric in \dt. The
last term will dominate over the second-to-last term since
$\dm\gg\Delta\Gamma$.
In this analysis, we assume $\Delta \Gamma$ and \CP violating effects
are negligibly small~\cite{PDG2002,cp}. We extract \dm, \fpfz the
ratio of \ups branching fractions to $B^+ B^-$ and $B^0 \bar{B}^0$,
and the \CPT violation parameters $\Re(\cos \theta)$ and $\Im (\cos
\theta)$. Our previous determination of $\Delta m_d$ using dilepton
events from 5.9 \invfb~\cite{PRLDL} treated \fpfz as a fixed
parameter. Otherwise the results reported here use the same analysis
method and include the earlier data and, therefore, supersede the
previous values.

\section{Event Selection}
The Belle detector, which consists of a silicon vertex detector (SVD),
a central drift chamber (CDC), aerogel \v{C}erenkov counters (ACC),
time-of-flight counters (TOF), an electromagnetic calorimeter (ECL),
and a muon and $K_L$ detector (KLM), is described in detail
elsewhere~\cite{NIM}.
For electron identification, we use position, cluster energy, and
shower shape in the ECL, combined with track momentum and \dedx in
the CDC and hits in the ACC.  For muon identification, we extrapolate
the CDC track to the KLM and compare the measured range and transverse
deviation in the KLM with the expected values.

\subsection{Lepton Selection}
The efficiencies for identifying leptons are determined from
two-photon process data samples: $e^+ e^- \to e^+ e^- e^+ e^-$ for
electrons and $e^+ e^- \to e^+ e^- \mu^+ \mu^-$ for muons. For both
cases, the possible degradation of efficiency due to nearby tracks
that are not present in QED events but must be considered in hadronic
events is examined using special hadronic event data samples that
contain embedded Monte Carlo (MC) lepton tracks.

We determine the probabilities for misidentifying hadrons as leptons
using data samples of $K_S \to \pi^+\pi^-$ decays for pions, $\phi \to
K^+ K^-$ decays for kaons, and $\Lambda \to p \pi^-$ decays for
protons. For tracks in the kinematic region of the dilepton event
selection, the identification efficiencies are 92.6\% for electrons
and 87.0\% for muons. About 0.1\% of pions and kaons, 0.2\% of protons
and 1.2\% of antiprotons are misidentified as electrons. About 1\% of
pions and kaons, and 0.2\% of protons and antiprotons are
misidentified as muons.

\subsection{Hadronic Event Selection}
Hadronic events are selected from a data set corresponding to
29.4~\invfb at the \ups resonance and 3.0~\invfb at an energy 60~\mev
below the peak. Hadronic events are required to have at least five
tracks, an event vertex with radial and $z$-coordinates (where the $z$
axis passes through the nominal interaction point, and is antiparallel
to the positron beam) within 1.5~cm and 3.5~cm respectively of the
nominal interaction point (IP), a total reconstructed centre-of-mass
(CM) energy greater than $0.5~W$ ($W$ is the \ups CM energy), a $z$
component of the net reconstructed CM momentum less than $0.3~W/c$, a
total ECL energy deposit between $0.025~W$ and $0.9~W$, and a ratio
$R_2$ of the second and zeroth Fox-Wolfram moments~\cite{Fox-Wolfram}
less than 0.7.

\subsection{Dilepton Event Selection}
Lepton candidates are selected from charged tracks that have a
distance of closest approach to the run-dependent IP less than 0.05~cm
radially ($dr_{\mathrm{IP}}$) and 2.0~cm in $z$ ($dz_{\mathrm{IP}}$).
At least one $r$-$\phi$ and two $z$ hits are required in the SVD. To
eliminate electrons from $\gamma \to e^+ e^-$ conversions, electron
candidates are paired with all other oppositely charged tracks and the
invariant mass (assuming the electron mass hypothesis) $M_{e^+e^-}$ is
calculated. If $M_{e^+e^-} < 100~\mevcc$, the candidate electron is
rejected. If a hadronic event contains more than two lepton
candidates, we use the two candidates with the highest CM momenta.

The CM momentum of each lepton is required to be in the range
$1.1~\gevc < p^\ast < 2.3~\gevc$. The lower cut reduces contributions
from secondary (charm) decay. The upper cut reduces the contribution
from non-$B\bar{B}$ continuum events. The angle of each lepton track
with respect to the $z$ axis in the laboratory frame must satisfy
30$^\circ < \theta_{\mathrm{lab}} <$ 135$^\circ$. This cut rejects
tracks with large angles of incidence in the SVD and hence provides
better $z$ vertex resolution. In addition, these cuts remove lepton
candidates whose particle identification is performed using the endcap
KLM or ECL, where the performance is not as good as that of the barrel
sections. Events that contain one or more \jpsi candidates are
rejected. We calculate the invariant mass of each candidate lepton
with each oppositely charged track (assuming the correct lepton mass
hypothesis). If the invariant mass falls into the \jpsi region,
defined as $-0.15~\gevcc < ( M_{e^+ e^-} - M_{\jpsi})< 0.05~\gevcc$,
$-0.05~\gevcc < ( M_{\mu^+ \mu^-} - M_{\jpsi})< 0.05~\gevcc$, the
candidate event is rejected. The looser lower cut for the electron
pair invariant mass is to reject \jpsi mesons whose calculated mass is
low due to bremsstrahlung of the daughter electron(s). The opening
angle of the two leptons in the CM frame $\theta^\ast_{\ell\ell}$, is
required to satisfy $ -0.8 < \cos \theta^\ast_{\ell \ell} < 0.95$ to
reduce jet-like continuum events, SS events where the two
reconstructed tracks originate from the from the same particle and
events with a primary lepton and a secondary lepton originating from
the same \B meson. After all cuts are applied, we obtain 49838 SS and
230881 OS events. The numbers of selected dilepton events are
summarised in table~\ref{eventtable}.
\begin{table}[!htbp]
  \begin{center}
    \caption{Summary of dilepton events after all cuts. The numbers in
      the off-resonance columns are scaled using the luminosity ratio
      and represent the contribution included in the on-resonance
      data.}
    \label{eventtable}
    \begin{tabular}{|c|rr|rr|}
      \hline \hline                                                                   
      Lepton flavours  &\multicolumn{2}{c|}{On-resonance}
      &\multicolumn{2}{c|}{Off-resonance}\\
      \hline \hline
      &  \multicolumn{1}{c}{SS}
      &  \multicolumn{1}{c|}{OS}
      &  \multicolumn{1}{c}{SS}
      &  \multicolumn{1}{c|}{OS} \\
      \hline
      $e e$         &9877     &52141     &107.4     &1513.3  \\
      $\mu \mu$     &15503    &65435     &1464.4   &4451.9  \\
      $e \mu$       &24458    &113305    &976.3    &4403.1  \\
      \hline
      total         &49838    &230881    &2548.1   &10368.2 \\
      \hline \hline
    \end{tabular}
  \end{center}
\end{table}

\section{\dz Determination}
The $z$-coordinate of each \B meson decay vertex is determined from
the intersection of the lepton track with the run-dependent profile of
the IP smeared in the $r$-$\phi$ plane by 21 $\mu$m to account for the
transverse \B flight length. We define the difference between the
$z$-coordinates of the two leptons as $ \dz = z(\ell^+) - z(\ell^-)$
for OS events and $ \dz = |z(\ell^\pm) - z(\ell^\pm)|$ for SS events.
\dz is related to the proper-time difference by $\dz \simeq
c\beta\gamma\dt$. The Lorentz boost factor of the $e^{+}e^{-}$ CM
frame at KEKB is $\beta\gamma = 0.425$~\cite{KEKB}.

The observed \dz distributions have contributions from ``signal''
defined as events where both leptons are primary leptons from
semileptonic decays of \B mesons, and ``background'' where at least
one lepton is secondary or fake, or the event is from the continuum.
The contributions from ``signal'' are theoretically well-defined
distributions convolved with the detector response function, which
describes the difference between the true \dz and the measured \dz. In
order to estimate the detector response function, we employ \jpsi
decays, where the true \dz is equal to zero, and whose measured \dz
distribution, after the contribution of background is subtracted,
yields the response function.

\begin{figure}[!htbp]
\begin{center}
\includegraphics[width=0.98 \columnwidth,height=!]{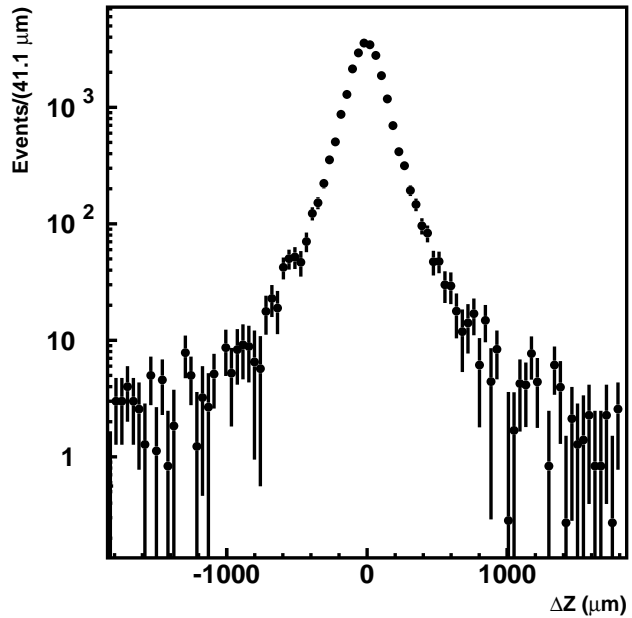}
\caption{Detector response function determined from the \jpsi data. }
\label{response}
\end{center}
\end{figure}

Candidate \jpsi events are selected using the same criteria as the
dilepton sample except the \jpsi veto and the cuts on $\cos
\theta^\ast_{\ell\ell}$ are not applied. We define the \jpsi signal
region as $3.00~\gevcc < M_{e^+ e^-} < 3.14~\gevcc$ and $3.05~\gevcc <
M_{\mu^+ \mu^-} < 3.14~\gevcc$, and the sideband region as
$3.18~\gevcc < M_{\ell^+ \ell^-} < 3.50~\gevcc$ for both electrons and
muons. A linear function is fitted to the sideband region of the mass
distribution and extrapolated to the signal region. Using the fit
result, we scale the \dz distribution in the sideband region to the
background underneath the peak in the signal region, and subtract it
from the signal region \dz distribution. Figure~\ref{response} shows
the resulting detector response function. It has $\mathrm{RMS} =
186~\mu\mathrm{m}$ in the range $|\dz | < 1850~\mu\mathrm{m}$. We use
this histogram as a lookup table in the analysis.

\section{Fitting}       
The mixing parameter \dm and other parameters are extracted
by simultaneously fitting the \dz distributions of SS
and OS events to the sum of contributions from all known
signal and background sources. We use a binned maximum likelihood
method. The background \dz distributions are obtained from
MC and used in the fits in the form of lookup tables.

\subsection{Signal \dz Distributions}
The signal from the neutral \B meson pairs originates either from $B^0
\bar{B}^0$ (unm) or from $B^0 B^0$ and $\bar{B}^0 \bar{B}^0$ (mix). If
\CPT is assumed to be conserved, the signal distributions are
expressed by
\begin{eqnarray}
  P^{\mathrm{unm}} =&  
   N_{4S} \, f_0 \, b_0^2 \, \epsilon_{\ell\ell}^{\mathrm{unm}} 
   \frac{e^{-\frac{|\dt|}{\tau_{B^0}}}}{4\tau_{B^0}} 
    [ 1 + \cos(\Delta m_d \dt) ]\;  
\nonumber \\
  P^{\mathrm{mix\;}} =&
   N_{4S} \, f_0 \, b_0^2 \, \epsilon_{\ell\ell}^{\mathrm{mix\;\,}} 
   \frac{e^{-\frac{|\dt|}{\tau_{B^0}}}}{4\tau_{B^0}} 
   [ 1 - \cos(\Delta m_d \dt) ] . 
\label{eq:defaultpdf}
\end{eqnarray}
If the possibility of \CPT violation is included, these expressions become, 
\begin{eqnarray}
  P^{\mathrm{unm}} =& 
  N_{4S} \, f_0 \, b_0^2 \, \epsilon_{\ell\ell}^{\mathrm{unm}} 
  \frac{e^{-\frac{|\dt|}{\tau_{B^0}}}}{4\tau_{B^0}} 
  [( 1 - |\cos \theta|^2) \cos(\dm \dt)   \nonumber \\
  &+ 1 + |\cos \theta|^2 - 2 \Im(\cos \theta) \sin(\dm \dt)]
  \nonumber \\
  P^{\mathrm{mix\;}} =&
  N_{4S} \, f_0 \, b_0^2 \, \epsilon_{\ell\ell}^{\mathrm{mix}} 
  \frac{e^{-\frac{|\dt|}{\tau_{B^0}}}}{4\tau_{B^0}}
  | \sin \theta |^2 [1-\cos(\Delta m_d \dt)]. \,\,\,\,\,
  \label{eq:cptpdf}
\end{eqnarray}
The integrals of $P^{\mathrm{unm}}$ and $P^{\mathrm{mix}}$ give  
the time-integrated fraction of mixed events %in the dilepton events  
\begin{equation}
\chi_d = \frac{| \sin \theta |^2 x_d^2}
{| \sin \theta |^2 x_d^2 + 2 + x_d^2 + | \cos \theta |^2 x_d^2}
\label{eq:chi}
\end{equation}
where $x_d \equiv \tau_{B^0} \dm $. When \CPT is conserved, 
this becomes the more familiar expression $\chi_d = x_d^2/(2 + 2x_d^2)$.
The values of $|\cos \theta |^2$ and $|\sin \theta |^2$ are determined from
$\Re(\cos\theta)$ and $\Im(\cos\theta)$.

The signal distribution for charged \B meson pairs is the same
for both the \CPT conserving and \CPT violating cases and given by
\begin{equation}
  P^{\mathrm{chd}} = N_{4S} \, f_+ \, b_+^2  
       \epsilon_{\ell\ell}^{\mathrm{chd}} \,
           \frac{e^{-\frac{|\dt|}{\tau_{B^+}}}}{2\tau_{B^+}}. 
\label{eq:chdpdf} 
\end{equation}
In the equation above, $N_{4S}$ is the total number of \ups events,
$f_0$ and $f_+$ are the branching fractions of \ups to neutral and
charged \B pairs ($f_+ + f_0 = 1$), $b_0$ and $b_+$ are the
semileptonic branching fractions for neutral and charged \B mesons,
$\epsilon_{\ell\ell}$ with superscript are the efficiencies for
selecting dilepton events of charged (chd), unmixed (unm), and mixed
(mix) origins. The ratio $\epsilon_{\ell\ell}^{\mathrm{chd}}$:
$\epsilon_{\ell\ell}^{\mathrm{unm}}$:
$\epsilon_{\ell\ell}^{\mathrm{mix}}$ is determined from MC and is
fixed in the fit, because any detector effect that is not simulated
correctly should affect events with these origins equally. The \dz
distributions are obtained from these distributions (equations
\ref{eq:defaultpdf}, \ref{eq:cptpdf} and \ref{eq:chdpdf}) by
conversion from \dt and convolution with the emperical resolution
function described in the previous section.

\subsection{Background \dz Distributions}
The background \dz distributions are estimated using the MC. A
comparison of the \dz distribution of \jpsi data samples between the
data and MC shows that the data distribution is wider than the MC
distribution. A detailed study showed that after convolving the MC
distribution with a $\sigma = 50 \pm 18~\mu$m Gaussian, the
distributions compared favourably. We smear MC background
distributions in the same way to compensate for this discrepancy.

We categorise the backgrounds into eight types depending on their
sources: charged \B pairs, mixed and unmixed neutral \B pairs, and
continuum, each of them contributing to both SS and OS events.
To normalise the amount of continuum background, we use the
off-resonance data. This leaves seven parameters to normalise the
fractions of other backgrounds. The first of these is $\chi_d$ which
is varied in the fit as given by equation~\ref{eq:chi}; the remaining
six are associated with the efficiencies for selecting these
background events which are denoted as
\begin{equation}
\epsilon_{SS}^{\mathrm{chd}},~~\epsilon_{OS}^{\mathrm{chd}},
              ~~\epsilon_{SS}^{\mathrm{unm}},~~ 
\epsilon_{OS}^{\mathrm{unm}},~~\epsilon_{SS}^{\mathrm{mix}},
                    ~~\epsilon_{OS}^{\mathrm{mix}}.
\end{equation}
We combine these six types into two: correct-tag (CT) which is
associated with $\epsilon_{OS}^{\mathrm{chd}}$,
$\epsilon_{OS}^{\mathrm{unm}}$, $\epsilon_{SS}^{\mathrm{mix}}$, and
wrong-tag (WT) which is associated with
$\epsilon_{SS}^{\mathrm{chd}}$, $\epsilon_{SS}^{\mathrm{unm}}$,
$\epsilon_{OS}^{\mathrm{mix}}$. For both CT and WT backgrounds,
relative fractions of the three background types are fixed according
to the MC. We then determine the \dz distributions for CT and WT
backgrounds by adding the three corresponding contributions.
    
The shapes of the \dz background distributions and the normalisation
of the backgrounds from neutral \B events depend on $\Delta m_d$. To
account for this, we generated two samples of generic neutral \B MC
events, one with $\Delta m_d = 0.469~{\mathrm{ps^{-1}}}$ and one with
$\Delta m_d = 0.522~{\mathrm{ps^{-1}}}$. Background \dz distributions
for any value of $\Delta m_d$ are produced by linear interpolation
between these two MC data sets.

The MC simulation does not always reproduce the hadron showering
processes correctly in the kinematic region of interest. In order to
account for possible discrepancies, we obtain an overall correction
factor by using a special control data sample. In this sample events
are selected in the same way as the dilepton events, except that we
now require that exactly one lepton passes the lepton selection
criteria and the other track passes all selection criteria except for
the lepton identification requirements. The ratio of data to MC fake
rates binned in $\theta_\mathrm{lab}$ and $p$ is then applied to the
control sample to obtain the overall fake rates for the dilepton
analysis. From this method, the hadron misidentification probabilities
of the MC are increased by 6\% for muons and decreased by 5\% for
electrons.

\section{Results}
In the fit, we fix the parameters $\tau_{B^0} = 1.542 \pm
0.016~{\mathrm{ps}}$, and $\tau_{B^+}/\tau_{B^{0}} = 1.083 \pm
0.017$~\cite{PDG2002}, and we impose the constraint $b_+/b_0 =
\tau_{B^+}/\tau_{B^0}$. In the fit where \CPT is conserved,
we have a total of five parameters to be fitted. They are \dm, \fpfz,
two parameters which are related to the fractions of CT
and WT backgrounds and are expressed as $T_{CT} \equiv
b_{+}^2 \epsilon_{\ell\ell}^{\mathrm{chd}}/
\epsilon_{OS}^{\mathrm{chd}}$ and $T_{WT} \equiv b_{+}^2
\epsilon_{\ell\ell}^{\mathrm{chd}}/ \epsilon_{SS}^{\mathrm{chd}}$, and
an overall normalisation. In the search for \CPT violation, we add
two parameters $\Re(\cos \theta)$ and $\Im(\cos \theta)$. The results
of the fits are summarised in table~\ref{fitresult}.
\begin{table}[!htbp]
  \begin{center}
    \caption{Results of fits.}
    \label{fitresult}
    \begin{tabular}{|c|c|c|}
      \hline \hline
      Fitting method    &\CPT conserved       & Allow \CPT Violation \\
      \hline \hline
      $\Delta m_d~({\mathrm{ps^{-1}}})$ &$0.503 \pm 0.008$ & $0.503 \pm 0.008$ \\
      \hline
      $\fpfz$         &$ 1.01 \pm 0.03$  &$1.02 \pm 0.03$        \\
      \hline
      $\Re (\cos \theta)$&                  &$0.00 \pm 0.12$        \\
      \hline
      $\Im (\cos \theta)$&                  &$0.03 \pm 0.01$        \\
      \hline
      $T_{CT}$          &$94 \pm 6$        &$91 \pm 5$             \\
      \hline
      $T_{WT}$          &$140 \pm 2$       &$139 \pm 2$            \\
      \hline
      $\chi^2$          &139 ($ndf =86$)   &132 ($ndf = 84$)       \\
      \hline \hline
    \end{tabular}
  \end{center}
\end{table}

\begin{figure}[!htbp]
\begin{center}
\includegraphics[width=0.98 \columnwidth,height=!]{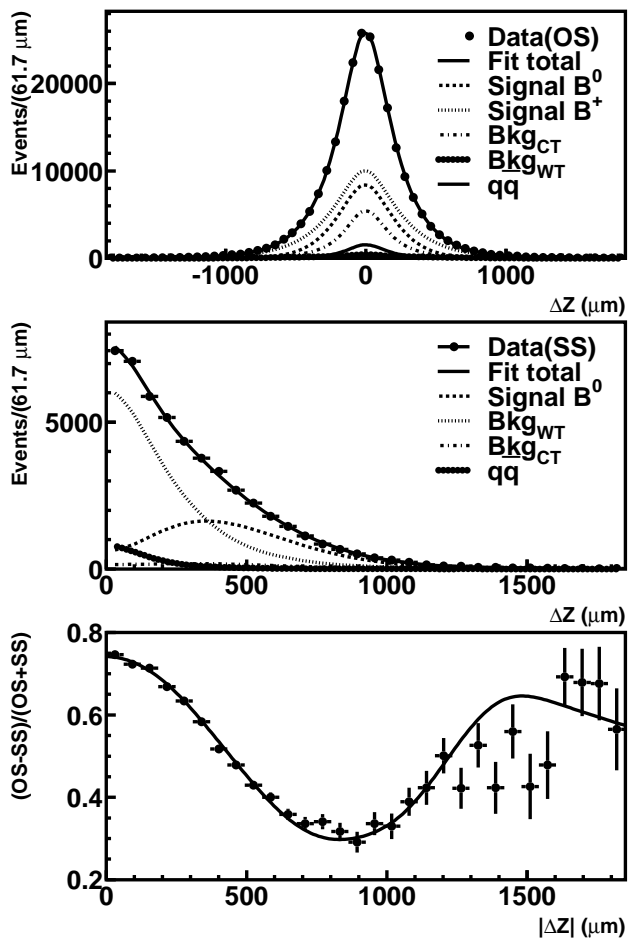}
\caption{Results of the simultaneous fit to the same-sign and
  opposite-sign \dz distributions assuming \CPT invariance.
  The upper two plots are for the opposite-sign and
  same-sign events, respectively. Signal and background obtained from
  the fit are also shown. The bottom plot is the OS-SS asymmetry with
  the fit result superimposed.}
\label{deltazasym}
\end{center}
\end{figure}

The values of \dm  and \fpfz do not vary in a significant
way when \CPT violation is included.  We use the result from the
\CPT conserving fit to obtain \dm and \fpfz.  The \dz
distributions for the SS and OS events and the resulting 
asymmetry are shown in Fig.~\ref{deltazasym},
together with the fit results.

Systematic errors in the determination of \dm and \fpfz include
contributions from uncertainties in the parameters that are fixed in
the fit: $\tau_{B^+}/\tau_{B^0}$, the extra $50~\mu$m smearing of the
background \dz distributions, $\tau_{B^0}$, inclusive $D$ meson
branching fractions $B \to D^+ X$ and $B \to D^0 X$. Also contributing
to the systematic errors are uncertainties in the continuum
contribution determined from off-resonance data, lepton
misidentification probabilities, detector response function, MC
statistics, background function linear interpolation end points, IP
profile smearing, the event selection criteria $\theta_{\mathrm{lab}}$
and $dr_{\mathrm{IP}}$, and the \dz fit range. Uncertainties of the
overall $z$ scale in the detector and the boost factor $\beta\gamma$
can also contribute to the systematic error.

The contribution from the detector response function is dominated by
the statistics of the \jpsi event sample. Biases due to approximating
the detector response function with the \dz distribution of \jpsi
events and the effect of the \B motion in the \ups frame also have
some contribution. We estimate the uncertainty of the detector
response function by comparing the results of fits to a full MC
simulation using three different types of response functions. These
are the \jpsi response function (extracted in the same way as in
data), a true \dz response function, and a response function based on
the proper time difference between the \B meson decays (for systematic
errors associated with the \B motion in the \ups frame).
The true \dz response function is constructed using signal events and
subtracting the true \dz from the measured \dz. The proper time
response function is constructed in the same manner, except the true
$\beta\gamma c \dt$ is subtracted.
Each of the three types of response functions is generated using 35
independent MC data sets, resulting in 105 response functions. The fit
results using each response function are then compared to extract the
systematic errors.
Additionally we vary the amount of background subtraction from the
\jpsi mass peak by $\pm 1~\sigma$, and repeat the fits.
The final detector response function systematic error is the combined
error resulting from the above methods.

To quantify the systematic error associated with the MC statistics,
the MC is divided into $n=20$ sets and the data is fitted using each
background distribution $P_i$, $(i=1...n)$. Both the shapes and the
efficiency ratios are independent for each fit. The fits were repeated
with 45 different pairs of end points for the \dm linear
interpolation. The RMS of the fit results was assigned as the error.
The systematic error for the IP constraint was estimated by varying
the smearing used to represent the transverse \B flight length by $\pm
10~\mu$m. The cuts on the variables \dz, $\theta_{\mathrm{lab}}$ and
$dr_{\mathrm{IP}}$ were varied in the region of their default values
and the changes in the fit results assigned as errors. The remaining
systematic errors were calculated by varying the default values by
$\pm 1~\sigma$, repeating the fits and assigning the differences as
errors. The contributions from the uncertanies in the $z$ scale and
$\beta\gamma$ were found to be negligible.

The errors are summarised in table~\ref{syserror}.
\begin{table}[!htbp]
  \begin{center}
    \caption{Systematic errors contributing to the \dm and \fpfz 
      measurements.}
    \label{syserror}    
    \begin{tabular}{lcc}\hline \hline
      Source & $\dm(\invps)$            & \fpfz   \\ \hline \hline
      $\tau_{B^+}/\tau_{B^0}$ ($\pm 0.017$)& $\pm 0.0053$     & $\pm 0.071$   \\
      Detector response function
      & $\pm 0.0047$     & $\pm 0.021$   \\
      Monte Carlo statistics  & $\pm 0.0036$     & $\pm 0.011$   \\
      $50~\mu$m smearing of bkg \dz ($\pm 18~\mu$m)
      & $\pm 0.0032$     & $\pm 0.015$    \\
      $\dz$ cut
      & $\pm 0.0030$ & $\pm 0.010$ \\
      $\theta_{\mathrm{lab}}$ cut
      &$\pm 0.0030$  & $\pm 0.030$ \\
      $\tau_{B^0}$    ($\pm 0.016~\mathrm{ps}$)
      & $\pm 0.0012$     & $\pm 0.002$   \\
      $dr_{\mathrm{IP}}$ cut
      & $\pm 0.0010$ & $\pm 0.003$ \\
      Continuum contribution   & $\pm 0.0008$     & $\pm 0.001$   \\
      $Br(B \to D X)$($D^0$:\,$\pm 4.7\%, D^+$:\,$\pm 7.9\%$)
      & $\pm 0.0007$     & $\pm 0.001$    \\
      Fake rate correction ($\mu$:\,$\pm 3\%$,\,$e$:\,$\pm 25\%$)
      & $\pm 0.0007$     & $\pm 0.002$     \\
      Linear interpolation
      & $\pm 0.0005$ & $\pm 0.020$ \\
      IP profile ($\pm 10 ~ \mu$m)
      & $\pm 0.0002$ & $\pm 0.005$ \\
      \hline
      Quadratic sum     & $\pm 0.0097$     & $\pm 0.086$   \\   
      \hline \hline
    \end{tabular}
  \end{center}
\end{table}

To reduce the backgrounds, the analysis was repeated with tighter
$p^\ast$ cuts, and separate fits were performed with $ee$, $e\mu$ and
$\mu\mu$ sub-samples. Deviations from the default results were all
consistent with statistical fluctuations. We repeated the fit
including the effects of $\Delta\Gamma/\Gamma=1\%$ and found the shift
in results to be negligible ($\approx 0.0001~\invps$).

The largest contributions to the systematic error for $\Im (\cos
\theta)$ come from the uncertainty in the extra $50~\mu$m smearing and
the $\theta_{\mathrm{lab}}$ selection criteria and amount to $\pm
0.03$. The systematic error for $\Re (\cos \theta)$ is dominated by the
uncertainty in the response function. We conservatively assign an
error of $\pm 0.01$.

In summary, we obtain    
\begin{eqnarray}
\Delta m_d &=& (0.503 \pm 0.008({\rm stat}) \pm 0.010({\rm sys}))
  ~{\mathrm{ps^{-1}}} \nonumber \\
\fpfz &=& 1.01 \pm 0.03({\rm stat}) \pm 0.09({\rm sys}) \nonumber \\
\Re (\cos \theta) &=& 0.00 \pm 0.12({\rm stat}) \pm 0.01({\rm sys}) 
                           \nonumber \\
\Im (\cos \theta) &=& 0.03 \pm 0.01({\rm stat}) \pm 0.03({\rm sys}). 
\end{eqnarray}
Using world averages for $\Delta m_d$, the neutral \B meson mass
and decay width~\cite{PDG2002}, these \CPT parameters imply~\cite{PRD}
the upper limits $| m_{B^0} - m_{\bar{B}^0} | /m_{B^0} < 1.16
\times 10^{-14}$ and $| \Gamma_{B^0} -\Gamma_{\bar{B}^0}
| / \Gamma_{B^0} < 0.11$ at 90\% C.L..

\section{Summary}
In summary, we have measured $\Delta m_d$ and the ratio of branching
fractions for \ups decay to $B^+ B^-$ and $B^0 \bar{B}^0$, \fpfz,
using inclusive dilepton events. The largest contributions to the
systematic error for the \dm measurement come from the response
function and the uncertainty in the \B meson lifetime ratio
$\tau_{B^+}/\tau_{B^0}$. Recent measurements~\cite{PDG2002}
of this lifetime ratio have significantly reduced this systematic
contribution since our first measurement~\cite{PRLDL}. The result of
this dilepton analysis $\Delta m_d = (0.503 \pm 0.008 \pm
0.010)~{\mathrm{ps^{-1}}}$ is in good agreement with the results from
other methods used by Belle~\cite{otherbelle} and the current world
average~\cite{PDG2002}. The error for the \fpfz measurement is
dominated by the uncertainly in $\tau_{B^+}/\tau_{B^0}$. We have also
obtained new limits on \CPT violation parameters.

\section*{Acknowledgements}
We wish to thank the KEKB accelerator group for the excellent
operation of the KEKB accelerator.
We acknowledge support from the Ministry of Education,
Culture, Sports, Science, and Technology of Japan
and the Japan Society for the Promotion of Science;
the Australian Research Council
and the Australian Department of Industry, Science and Resources;
the National Science Foundation of China under contract No.~10175071;
the Department of Science and Technology of India;
the BK21 program of the Ministry of Education of Korea
and the CHEP SRC program of the Korea Science and Engineering Foundation;
the Polish State Committee for Scientific Research
under contract No.~2P03B 17017;
the Ministry of Science and Technology of the Russian Federation;
the Ministry of Education, Science and Sport of the Republic of Slovenia;
the National Science Council and the Ministry of Education of Taiwan;
and the U.S.\ Department of Energy.
\\

\end{document}